\documentstyle[fl eqn,twoside]{article}
\def\be{\begin{equation}}
\def\ee{\end{equation}}
\def\bi{\bibitem}
\begin{document}
\title{Hamiltonian formulation of curvature squared action}
\author{Abhik Kumar Sanyal}
\maketitle
\noindent
\begin{center}
Dept. of Physics, Jangipur College, Murshidabad,
\noindent
India - 742213\\
\noindent
and\\
\noindent
Relativity and Cosmology Research Centre\\
\noindent
Dept. of Physics, Jadavpur University\\
\noindent
Calcutta - 700032, India\\
\noindent
e-mail : aks@juphys.ernet.in\\
\end{center}
\noindent
\date{}
\maketitle
\begin{abstract}
It has been observed earlier that, in principle, it is possible to obtain a 
quantum mechanical interpretation of higher order quantum cosmological 
models in the spatially homogeneous and isotropic background, if auxiliary 
variable required for the Hamiltonian formulation of the theory is chosen 
properly. It was suggested that for such a choice, it is required to get rid 
of all the total derivative terms from the action containing higher order 
curvature invariant terms, prior to the introduction of auxiliary variable. 
Here, the earlier work has been modified and it is shown that the action, 
$A=\beta\int~\sqrt{-g}~R^2~ d^4 x$ should be supplemented by a boundary term 
in the form  $\sigma=4\beta\int ~^{3}R~ K~ \sqrt{h}~d^{3}x$, where, $^3 R$, 
$K$ and $h$ are the Ricci scalar for three-space, trace of the extrinsic 
curvature and determinant of the metric on the three space respectively. The 
result has been tested in the background of homogeneous and anisotropic models 
and thus confirmed.
\end{abstract}
PACS .98.80.Hw,04.50.+h
\section{\bf{Introduction}}
Higher order theories were originally introduced to get rid of ultraviolet divergence appearing in the process of quantizing gravity \cite{w:t}. Stell  \cite{k:p} has shown that Einstein-Hilbert action being modified by the introduction of curvature squared term in the form, $\int d^4{x}\sqrt{-g}[AC_{ijkl}+BR+CR^2]$ is renormalizable in 4-dimensions, where, $A, B, C$ are the coupling constants. Further, such a theory has also been found to be asymptotically free \cite{t:f}. Unitarity of higher derivative quantum theory of gravity has been established by Tomboulis \cite{t:p}. In an attempt to explain the inflationary scenario without invoking phase transition in the very early universe, Starobinsky \cite {s:p} considered a field equation containing only geometric terms. Starobinsky and Schmidt \cite {ss:c} later have shown that the inflationary phase is an attractor in the neighbourhood of the solution of the fourth order gravity theory. Hawking and Luttrell \cite{h:n} on the other hand observed that under suitable conformal transformation, Einstein-Hilbert action together with a curvature squared term turns out to be equivalent to the Einstein-Hilbert action being coupled to a massive scalar field. Recently, the importance of considering $R^2$ term in the action has further been increased, as it has been observed that the $4$-dimensional Brane world \cite{r:s} effective action \cite{k:s} contains such term. There is even stronger motivation to consider higher order theory of gravity, which we discuss below.
\par
 In analogy with the particle quantum dynamics and the quantum field 
theory, Hartle-Hawking \cite {hh:p} suggested that the ground state wave 
function of the universe should be given by
\[
\psi_{0}[h_{ij}] = N \int~~\delta g_{\mu\nu} ~exp(-I_{E}~[g_{\mu\nu}]),
\]
along with the proposal that the sum should be taken over all compact four 
geometries. In the above expression $h_{ij}$ is the metric on the three space,
$I_{E} = -i~ A$ is the Euclidean counterpart of the Einstein-Hilbert action,- \[A = \int \frac{1}{16\pi G}~R\sqrt{-g}~d^4x,\] where, $R$ is the Ricci scalar and $g_{\mu\nu}$ is the metric of the four-space. This implies that the universe does not have any boundary in the Euclidean regime. The above functional integral over all compact four geometries bounded by a given three geometry is interpreted as the amplitude of a given three geometry to arise from a zero three geometry, where $\psi_{0}[h_{ij}]$ is finite. Thus the ground state is the amplitude of the universe to arise from nothing.
\par
This excellent prescription that attempted to give a probabilistic interpretation of quantum cosmology, for the first time, however, suffers from serious disease. The Euclidean form $I_{E}$ of the Einstein-Hilbert action $A$ is not positive definite and therefore the functional integral $\psi_{0}[h_{ij}]$ corresponding to the ground state wave function of the universe, as proposed by Hartle-Hawking runs into serious problem, since it diverges badly.
\par
A remedy to the above problem was suggested by Horowitz \cite {g:h}. To get a convergent integral for the grond state wave function $\psi_{0}[h_{ij}]$, he proposed an action in the form 

\[
S=\frac{1}{4} \int~~d^4 X\sqrt{-g}[A C_{ijkl}^2 +B (R- 4\Lambda)^2],
\]
where, $C_{ijkl}$ is the Weyl tensor, $\Lambda$ is the cosmological constant and $A, B$ are coupling constants. The above action which includes curvature squared term, is positive definite and in the weak energy limit it reduces to the Einstein-Hilbert action. It has there been shown that canonical quantization of the above action leads to a Schr\"{o}dinger like equation, where an internal variable acts as a time parameter.
\par
Perhaps the most attractive feature of higher order theory of gravity has been established in a series of recent publications \cite{a:m}, \cite{a:s}, \cite{a:n}. There, it has been shown that, under careful choice of auxiliary variable, required to cast such action in canonical form, the higher order theory of gravity, in homogeneous and isotropic minisuperspace model, under quantization, yields Schr\"{o}dinger like Wheeler-deWitt equation with excellant features. Firstly, one of the true degrees of freedom of the theory disentangles from the kinetic part of the canonical variables and behaves as the time parameter. Next, it gives birth to an effective Hamiltonian that turns out to be hermitian and as a result the continuity equation identifies the nature of space and time like variables in the Robertson-Walker minisuperspace model under consideration. Thus, a quantum mechanical idea of probability and current densities in quantum cosmology is established. Further, an effective potential emerges in the process, whose extremization yields vacuum Einstein's equation, which is ofcourse a desirable feature in the weak energy limit of higher order gravity theory. In view of all the results discussed above, we understand the need for modification of the Einstein-Hilbert action by the introduction of higher order curvature invariant terms. \par
Now, Lagrangian of a system usually involves field variables along with their first derivatives. This restriction is not required in classical context but somtimes becomes mandetory due to quantum mechanical requirements, such as positive definiteness of probability, existence of the vacuum etc. In the theory of gravitation under consideration, Lagrangian contains derivative of higher order, in the presence of higher order curvature invariant terms in the action. Thus, in such situations, in order to express the action in canonical form, that results in the Hamiltonian formulation of the theory, it is required to introduce auxiliary variable, in view of which the Lagrangian turns out to be the function of the field (along with the auxiliary one) variables and their first derivatives. For canonical formulation of such a theory, Ostrogradski's prescription \cite{o:m} is generally followed. It suggests to fix a spacelike surface $\Sigma$ in a space-time $(M, g_{\mu\nu})$, which is asymptotically flat. Canonical variables are the three metric $h_{ab}$, a quantity $Q_{ab}$ realated to the extrinsic curvature tensor and their conjugate momenta, $p^{ab}, P^{ab}$ respectively. Here, $Q_{ab}$ and $P^{ab}$ correspond to extra degree of freedom in higher derivative theory of gravity. The relations amongst $Q_{ab}, P^{ab}$, the space-time curvature $R$ and the extrinsic curvature $k_{ab}$ are
\[ Q^{ab} = 2k_{ab}\]
and
\[P_{ab}= -\beta\sqrt{h}h^{ab}~R\]
in the absence of electric part of the Weyl tensor. Thus the trace of the extra degree of freedom $Q$ is twice the trace of the extrinsic curvature and that of it's conjugate momentum $P$ is simply related to the scalar curvature and the phase space is spanned by $\{h, Q, p, P\}$. Hence, Ostrogradski's prescription \cite {o:m} requires as many extra degrees of freedoms as there are independent components of extrinsic curvature tensor.
\par
Boulware \cite{b:et} slightly modified the above prescription and suggested a proposal that the auxiliary variable may be chosen as $\pi_{ab}$, which is the momenta conjugate to the extrinsic curvature tensor $k_{ab}$, where, $\pi^{ab}=\int P^{ab}~d^3 x$, in the absence of the electric part of the Weyl tensor and ${h}= det(h_{ab})$. Both the above definitions are indeed confusing. It has been shown in some earlier works \cite{a:m}, \cite{a:s}, \cite{a:n} that, to get a well behaved quantum description of higher order theory of gravity, it is mandetory to get rid of certain boundary terms from the action. Canonical momenta have to be defined only after doing so, and as a result, $P_{ab}$ in Ostrogradski's prescription and $Q_{ab}$ in Boulware's proposal will not remain proportional to the curvature scalar $R$. This has been explicitely worked out in the following section. The quantum cosmological equation corresponding to the above action, presented by Horowitz \cite{g:h},
\[k_{ab}i\frac{\partial}{\partial{h_{ab}}}=G_{abcd}\frac{\partial}{\partial{k_{ab}}}\frac{\partial}{\partial{k_{cd}}}+A_{ab}\frac{\partial}{\partial{k_{ab}}}+V
\] 
has been constucted in view of the above definition of momentum. In the above $G_{abcd}$ is the metric of the superspace that depends on $h_{ab}$ and $A_{ab}$ is an artefact of the term $D_{a}D_{b}P^{ab}$, appearing in the Hamiltonian constraint equation. Now, Einstein-Hilbert action is supplemented by Gibbons-Hawking boundary term. Likewise, if it can be proved that action corresponding to higher order theory of gravity should be supplemented by an appropriate boundary term, then above quantum cosmological equation will change along with the Hamiltonian constraint equation and the definition of momentum.
\par 
While constructing a minisuperspace quantum cosmological model, Horowitz \cite{g:h}, suggested that the auxiliary variable can be chosen as the `negative of the derivative of the action with respect to the highest derivative of the field variable present in the action.' This prescription is again misleading, since one can then introduce auxiliary variable even in standard Einstein-Hilbart action, leading to completely wrong quantum dynamics. This has been shown in an earlier publication \cite{a:m}. One can get rid of such problem if one expresses the Einstein-Hilbert action along with Gibbons-Hawking term. This proves the importance of boundary term in the quantum domain. The problem is nontrivial when Einstein-Hilbert action is replaced or modified by higher order curvature invariant terms in the action, since there does not exist in general, surface terms in standard form for such theories.
\par
Our earlier proposal \cite{a:m}, \cite{a:s}, \cite{a:n} was to get rid of all removable total derivative terms from the action prior to the introduction of auxiliary variable. It worked nicely in the homogeneous and isotropic model. However, it has been observed here that, if we make a change of variable in the action by replacing the scale factor $a$ by $\sqrt{2z}$ then the total derivative terms existing in the action are different. This means total derivative terms are different under different choice of variables. So, our earlier suggestion of removal of all total derivative terms from the action prior to the introduction of auxiliary variable is not sensible enough. This implies that one has to suggest exactly the total derivative terms that should be eliminated from the action prior to the introduction of the auxiliary variable. This has been done in the present work. For an acid test of the present suggestion we extend our work in some of the homogeneous-anisotropic models. It has been observed that Boulare's prescription and our earlier suggestion raise lot of problems in finding even the classical field equations, when applied in anisotropic models, that correspond to more than one field variable in the action. Question is, which of the total derivative terms should be removed, since one of such terms can be expressed in terms of the other. Next question is, how many auxiliary variables should be introduced. In the present work we have attempted to give answer to the problems raised above. For this purpose we have considered an action containing only curvature squared term and studied both the isotropic and a class of anisotropic models.
\par
In the following section, we have studied some problems regarding the definition of momentum that has been discussed above. In sections 3 and 4, we explained the need for a boundary term to exhibit correct quantum description of a theory. In sections 5 and 6 we have been able to find an appropriate boundary term that would supplement curvature squared action, and have made a comparison with our earlier result. Sections 7 and 8 have been devoted to explain the problems arising in anisotropic model along with the remedy for such problems. In a nutshell, we have been able to prove that curvature squared action should essentially be supplemented by an appropriate boundary term.
\section{\bf{Problems regarding definition of momenta in $R^2$ Gravity}}
Hamiltonian formulation of a theory requires to express the action in canonical form. It has been pointed out in the introduction that, for the Hamiltonian formulation of an action containing curvature squared term in the form,
\be
A=\beta\int~{d^4x}\sqrt{-g}~R^2,
\ee
the corresponding Lagrangian in a minisuperspace model involves field variables along with their first derivatives and second derivatives. So, in order to cast the action in canonical form that demands the corresponding Lagrangian be expressed in terms of the field variables along with their first derivatives only, one requires to introduce auxiliary variables. Such auxiliary variables, should not be chosen in an adhoc manner. It has been observed that any form of auxiliary variable produces correct classical field equations in homogeneous and isotropic model. However, these have not so far been tested in anisotropic models. Further, it is not enough to get the correct classical field equations only, rather one should also be able to produce correct and well behaved, quantum description of the system under consideration. By the above statement we mean to find a variable, if it at all exists, with respect to which the Hamiltonian (it is the effective Hamiltonian in the present context) would turn out to be hermitian. With such motivation, canonical variables, as discussed in the introduction are usually chosen along the line of Ostrogrdski \cite{o:m} as the three metric $h_{ab}$, and the extrinsic curvature $k_{ab}$ on the compact three manifold $\Sigma$ along with their congugate momenta $p^{ab}$ and $\pi^{ab}$ respectively. Here, $p^{ab}$ contains third temporal derivative of the field variable and $\pi^{ab}$ is known to have the form $\pi^{ab}=\int P^{ab}~d^3 x$, While, $P^{ab}=
8\beta \sqrt{h}~h^{ab}~^4 R$, 
in the absence of the electric part of the Weyl tensor with, ${h}= det(h_{ab})$. For Hamiltonian formulation of such a theory Boulware \cite{b:et} chose $\pi^{ab}$ as the auxiliary variable and $k_{ab}$ as the conjugate momentum. So, according to Boulware's prescription, Hamiltonian formulation of an action containing higher order curvature invariant term  requires introduction of auxiliary variable, 
\be
Q^{ab}=\pi^{ab} = 8\beta\int~\sqrt{h}~ h^{ab}~^4 R~d^3 x
\ee
which would follow from the action $A$, if it is defined as 
\be
Q^{ab}=\pi^{ab} = \frac{\partial A}{\partial ({\frac{\partial k_{ab}}{\partial t}})}
\ee
\par 
To check whether these two definitions given in equations (2) and (3) are at par, we consider homogeneous and isotropic Robertson-Walker metric in the form
\be
ds^2=-dt^2+a(t)^2(d\chi^2+f^2(\chi)(d\theta^2+sin^2\theta~ d\phi^2)),
\ee
for which
\[
^4R=6(\frac{\ddot a}{a}+\frac{\dot a^2}{a^2}+\frac{k}{a^2})
\] 
\be
k_{11}=k_{22}=k_{33}=-a\dot{a},
\ee
where, $k$ is the curvature parameter. Now, in view of (2),
\be
Q^{11}=Q^{22}=Q^{33}=96\pi ^2 \beta (\frac{\ddot a}{a}+\frac{\dot a^2}{a^2}+\frac{k}{a^2}){a}
\ee 
Now the action (1) in view of (5) is,
\be
A=72\pi ^2 \beta\int~~(\frac{\ddot a}{a}+\frac{\dot a^2}{a^2}+\frac{k}{a^2})^2~ a^3 dt
\ee
It should be noted that if one varies the action with respect to the highest derivative ($\ddot a$) present in the action, as suggested by Horowitz \cite{g:h} and followed by pollock \cite{p:n}, Sanyal and Modak \cite{a:m} and Sanyal \cite{a:s}, one gets,
\be
\frac{\partial A}{\partial \ddot a}=144\pi ^2 \beta~ (\frac{\ddot a}{a}+\frac{\dot a^2}{a^2}+\frac{k}{a^2})a^2 
\ee
which does not resemble with the one given by equation (6) for the simple reason that the auxiliary variable here is the canonical momentum conjugate to the variable $\dot a$, which is different from $k_{ab}=-a\dot a$. In order to obtain $\pi ^{ab} = Q^{ab}$ through the definition given by (3) one has to choose a variable $z=\frac{a^2}{2}$ such that
\be
\dot z = a\dot a=-k_{aa}
\ee
The action (7), in view of this variable is expressed as, 
\be
A=36\sqrt 2 \pi ^2 \beta \int~~(\frac{\ddot z +k}{z})^2~ z^\frac{3}{2}~ dt
\ee
Therefore 
\be
Q= \pi^{aa}=\frac{\partial A}{\partial \ddot z}=72\sqrt{2}~\pi ^2 \beta \frac{\ddot z +k}{\sqrt z}=144\pi ^2\beta (\frac{\ddot a}{a}+\frac{\dot a^2}{a^2}+\frac{k}{a^2})a
\ee
Thus we see that $Q$'s given by (6) and (11) have the same form, while they differ only by a constant factor that can be arranged simply by changing the pre factor to $12$ instead of $8$, in the definition of $P^{ab}$ (and hence $\pi ^{ab}= Q^{ab}$ given in (2)). Thus,
\be
Q^{ab}=\pi^{ab} = 12\beta\int~~\sqrt{h}~ h^{ab}~^4R~d^3 x
\ee 
However, if one agrees with the fact that removable total derivative terms should be eliminated from the action prior to the definition of auxiliary variable, ie. action (1) should be supplemented by some boundary term, as suggested in our earlier works \cite{a:m}, \cite{a:s}, \cite{a:n}, then action (10) takes the following form,
\be
A= 36\sqrt 2 \pi^2 \beta \int (\frac{\ddot z^2}{\sqrt z}+k\frac{\dot z^2}{z\sqrt z}+\frac{k^2}{\sqrt z})dt+72\sqrt 2 \pi^2 \beta k\frac{\dot z}{\sqrt z}~~.
\ee
In view of which, 
\be 
Q=\pi^{aa}=72\sqrt 2 \pi^2 \beta \frac{\ddot z}{\sqrt z}= 144\pi^2 \beta (\frac{\ddot a}{a}+\frac{\dot a^2}{a^2})a,
\ee
does not match with the definition (12). Hence we find that the definition of momentum given by (2) or (12) is vulnarable and should not be taken seriously.
\par
Further, we can see that these two definitions given by (3) and either (2) or (12) do not agree in homogeneous-anisotropic models also. The Homogeneous and anisotropic Kantowski-Sachs, axially symmetric Bianchi-1 and Bianchi-111 minisuperspace models, can be expressed together in the following form,
\be
ds^2=-dt^2+a^2(t)dr^2+b^2(t)(d\theta ^2+f_{k} ^2(\theta)d\phi ^2),
\ee 
where, $k$ is the curvature index of the $2$-dimensional surface $d\theta ^2+f_{k} ^2(\theta)d\phi ^2$ and,
\[f_{k}~=sin\theta ,\] 
\[~~~~=\theta\]
\[~~~~=sinh\theta ,
\]
correspond to Kantowski-Sachs metric with positive spatial curvature, $k=1$, 
 Bianchi-1, with zero spatial curvature $k=0$ and Bianchi-111 metric, with negative spatial curvature $k=-1$ respectively. Corresponding Ricci scalar and extrinsic curvatures are given by,
\[
^4R=2(\frac{\ddot a}{a}+2\frac{\ddot b}{b}+2\frac{\dot a\dot 
b}{ab}+\frac{\dot b^2}{b^2}+\frac{k}{b^2}).
\]
\be
k_{11}=-a\dot a,~~~k_{22}=k_{33}=-b\dot b
\ee
Hence, In view of the definition (12),
\be
\pi^{11}=Q=48\pi \beta \frac{b^2}{a}~^4 R,~~\pi^{22}=\pi^{33}=q=48\pi \beta a~^4 R
\ee
So, from the prescription \cite{b:et}, it is clear that a pair of auxiliary variables $Q$ and $q$ is required to cast action (1) in canonical form in the anisotropic model under consideration. Now under the choice of the variables $x=\frac{a^2}{2}$ and $y=\frac{b^2}{2}$ such that,
\be
\dot x=a\dot a=-k_{11},~~~\dot y=b\dot b=-k_{22}=-k_{33}
\ee
and in view of (16), the action (1) takes the following form,
\be
A=8\sqrt{2}~\pi \beta\int~\sqrt{x}~y[\frac{\ddot x}{x}+2\frac{\ddot y}{y}-\frac{1}{2}(\frac{\dot x}{x}-\frac{\dot y}{y})^2+\frac{k}{y}]^2~dt
\ee
Hence, the auxiliary variables emerging from the canonical definition (3) are,
\[
Q=\frac{\partial A}{\partial \ddot x}=16\sqrt{2}~\pi\beta\frac{y}{\sqrt{x}}~^4 R=16\pi\beta\frac{b^2}{a}~^4 R
\]
\be
q=\frac{\partial A}{\partial \ddot y}=32\sqrt{2}~\pi\beta\sqrt{x}~^4 R=32\pi\beta a~^4 R
\ee
So, we observe that the auxiliary variables obtained in (17) and (20), in view of the two definitions (2) or (12) and (3) respectively, do not match. Further, as pointed out earlier, if one agrees that all removable total derivative terms should be removed from the action (1) prior to the definition of auxiliary variable, then after doing so, we have,
\[
A=8\sqrt 2 \pi \beta \int y\sqrt x [(\frac{\ddot x}{x}+2\frac{\ddot y}{y})^2 +2\frac{\dot x \dot y}{xy}(\frac{\ddot x}{x}+2\frac{\ddot y}{y})-\frac{\ddot x \dot y^2}{xy^2}-2\frac{\dot x^2 \ddot y}{x^2 y}-\frac{7\dot x^4}{12x^4}-\frac{13\dot y^4}{12y^4}-\frac{2\dot x^3 \dot y}{3x^3 y}-\frac{2\dot x \dot y^3}{3xy^3}+\]
\be
\frac{3\dot x^2}{2x^2}+\frac{\dot y^2}{y^2}+\frac{k}{y}(3\frac{\dot y^2}{y^2}+\frac{k}{y})]dt+8\sqrt 2 \pi \beta y\sqrt x[-\frac{1}{3}(\frac{\dot x^3}{x^3}+2\frac{\dot y^3}{y^3})+2\frac{k}{y}(\frac{\dot x}{x}+2\frac{\dot y}{y})]
\ee
Hence,
\[
Q=\pi^{11}=\frac{\partial A}{\partial{\ddot x}}=8\sqrt 2 \pi \beta \frac{y}{\sqrt x}[\frac{\ddot x}{x}+2\frac{\ddot y}{y}+\frac{\dot x \dot y}{xy}-\frac{1}{2}\frac{\dot y^2}{y^2}]
\]
\be
q=\pi^{22}=\pi^{33}=\frac{\partial A}{\partial {\ddot y}}=32\sqrt 2 \pi \beta \sqrt x [\frac{\dot x}{x}+2\frac{\ddot y}{y}+\frac{\dot x \dot y}{xy}-\frac{1}{2}\frac{\dot x^2}{x^2}]
\ee
Therefore it is noticed that different momenta and hence the auxiliary variables are revealed from the same action. Since, definition (3) is the canonical definition of momenta, therefore, the definition of $\pi^{ab}$ given by (2) or (12) is vulnerable and should be rejected. It should be emphasized at this stage that any attempt to quantize gravity starting from such a definition of momentum given by (2) or (12) is bound to produce wrong result. However, even after rejecting the definition (2) or (12), we are not at all free from ambiguity, since, definition (3) further, requires clarification, depending on whether action $A$ should be supplemented by some boundary term or not, and if it does, it furthar depends on the specific form the boundary term. In the following toy model we show that for a correct and well behaved quantum dynamics, any action indeed should be supplemented by a boundary term.
\section{\bf{The impotance of boundary term in quantum theory - a toy model }}
Let us take an action in the following form,
\[S=\int [\frac{1}{2}m\dot x^2+k\dot x-V(x)]dt .\]
It is seen that in the above action we have deliberately kept a term $k\dot x$, which is essentially a removable total derivative term. It is not difficult to see that the classical equation of motion is not affected any way in the presence of such a term. However, the definition of momentum is changed, ( which is now $p_{x}=m\dot x + k$ ), in view of which Hamiltonian changes considerabely,
\[H=\frac{p_{x}^2}{2m}- k\frac{p_{x}}{m} - \frac{m k^2}{2} + k^2 + V.\]
Such an Hamiltonian leads to a Schr\"{o}dinger equation that suffers from the problem of uniterity due to the presence of the second term which is linear in $p_{x}$. This simple situation teaches us the lesson that every action should be supplemented by an appropriate boundary term. It has been shown earlier \cite{a:m} that if we do'nt supplement Einstein-Hilbert action by Gibbon's-Hawking surface term then it's Hamiltonian formulation requires introduction of an auxiliary variable, which finally leads to wrong Wheeler-deWitt equation. Such arguments are convincing enough to make us believe that action containing higher-order curvature invariant terms must also be supplemented by a suitable boundary term. However, the requirement of such a boundary term is much apparent in the process of canonical quantization of $R^2$ theory of gravity via Boulware's prescription.  
\section{\bf{Canonical quantization of $R^2$ theory of gravity via Boulware's prescription - need for a boundary term}}
Before trying to find an appropriate boundary term that should supplement an action containing curvature squared ($R^2$) term given in the form (1), let us first observe the outcome of Boulware's prescription in homogeneous and isotropic model, if action (1) is not supplemented by a boundary term. As pointed out earlier, such prescription requires a set of phase space variables \{$h_{ab}, \pi^{ab}, p^{ab}, k_{ab}$\}. In this prescription, $h_{ab}$ and $\pi^{ab}$ are taken as the canonical vaiables whose corresponding momenta are  $p^{ab}$ and $k_{ab}$ respectively. For such a prescription (10) is the action and (11) is the auxiliary variable. For the corresponding Hamiltonian formulation, let us first express the action (10) in the following canonical form,
\be
A=\int~~[-\dot Q\dot z-\frac{Q^2 \sqrt{z}}{4M}+kQ]dt~+~Q\dot z,
\ee
where, $M=36\sqrt 2 ~\pi ^2 \beta$. Action (23) is canonical since the Hessian determinant $\Sigma\frac{\partial^2 L}{\partial \dot{q_{i}}\partial \dot{q_{j}}}$ is nonzero. In view of the above action, we observe that,
\be
p_{Q}=\frac{\partial L}{\partial \dot Q}=-\dot z = -a\dot a =k_{ab}.
\ee
Hence, Boulware's prescription is followed in the sense that, momentum, $\pi^{ab}= Q$ has been considered to be the field variable, while, $k_{ab}=\frac{\partial L}{\partial \dot Q}$, has been taken as the conjugate momentum. It is easy to see that the same definition of the auxiliary variable $Q$ is obtained from the above action along with the classical field equations. It is to be noted that finally we are left with a surface term $Q\dot z$, that can be expressed in the following form,
\be
\Sigma=4\beta\int~\sqrt{h}~K~^4 R~ d^3x,
\ee
where, $K$ is the trace of the extrinsic curvature. The Hamiltonian is given by,
\be
H=-p_{Q}~p_{z}+\frac{\sqrt z}{4M}Q^2-kQ
\ee
However if one is now interested in canonical quantization of the theory, it should be done with respect to the basic variables $h_{ab}, k^{ab}$ instead of $h_{ab}, Q$. Hence, we choose $\dot z = x$, which is essentially $-k_{ab}$, and replace $p_{Q}$ by $-x$ and $Q$ by $p_{x}$. As a consequence, 
\be
H=x p_{z}+\frac{\sqrt z}{4M}p_{x}^2-k p_{x}
\ee
The corresponding Wheeler-de-Witt equation is 
\be
i\hbar \frac{\partial\psi}{\partial z}=-\frac{\hbar ^2}{4M}\frac{\sqrt z}{x}\frac{\partial^2 \psi}{\partial x^2}+i\hbar \frac{k}{x}\frac{\partial\psi}{\partial x}.
\ee
The interesting feature to be noted is that, here we do not require to take into account any factor ordering ambiguity, since $x$ and $p_{z}$ appearing in the first term on the right hand side of the Hamiltonian (27) are independent variables, so also $z$ and $p_{z}$. Above equation can be expressed in the following form,
\be
i\hbar \frac{\partial \psi}{\partial z}=\hat{H_{0}}\psi
\ee
where, $\hat{H_{0}}$ is the effective Hamiltonian. Though this equation looks like Schr\"{o}dinger equation, it suffers from the problem that the effective Hamiltonian is not hermitian - the same problem that we have faced in the toy model above in section (3). Since, we have already been able to produce a hermitian effective Hamiltonian \cite {a:m} for the same situation under consideration, therefore we claim that equation (28) or (29) do not yield the correct quantum description of the theory.  Further, it has been pointed out earlier that unless an action is supplementad by an appropriate boundary term, it is always possible to introduce auxiliary variable, even in situations, where it is not required. In particular, It has been shown \cite{a:m} that if Einstein-Hilbert action is not supplemented by Gibbons-Hawking term then it is possible to introduce auxiliary variable, that keeps classical field equations unchanged while it produces wrong Wheeler-deWitt equation. Later, the same has been shown in the context of induced theory of gravity \cite{a:s}. This implies that boundary term plays a crucial role in quatum domain and every theory should be supplemented by an appropriate boundary term. Further, in the series of our earlier works \cite{a:m}, \cite{a:s}, it has been particularly pointed out that, if one removes all the removable total derivative terms prior to the introduction of auxiliary variable then the Schr\"{o}dinger equation corresponding to an action containing higher order terms like $R^2, R_{\mu\nu}R^{\mu\nu}$ in the homogeneous and isotropic model, exhibits some excellent features, viz., the effective Hamiltonian is hermitian, as a result of which quantum mechanical probabilistic interpretation of the theory is naturally revealed. Further, extremization of the effective potential gives classical vacuum field equation. In view of all these results we believe that action (1) must be supplemented by an appropriate boundary term and in search of that let us first remove all removable total derivative terms from action (10). 
\section{\bf{Removing boundary term in Boulware's prescription}}
In search of an appropriate boundary term for an action containing curvature squared term $R^2$, given in (1), let us start with action (10) and get rid of all the total derivative terms prior to the introduction of auxiliary variable. Thus we get action (13),
\[
A=M\int~[\frac{\ddot z^2}{\sqrt z}+k\frac{\dot z^2}{z\sqrt z}+\frac{k^2}{\sqrt z}]dt+2M k\frac{\dot z}{\sqrt z}.
\]
The boundary term that supplements the above action can be expressed in the form
\be
\sigma=4\beta\int~\sqrt h~K~^3 R~ d^3x
\ee
If we again proceed as before with $Q=\frac{\partial A}{\partial \ddot z}=2M \frac{\ddot z}{\sqrt z}$, given in (14), the action can be written in the following canonical form
\be
A=\int~[-\dot Q\dot z+Mk\frac{\dot z^2}{z\sqrt z}-\frac{\sqrt z}{4M}Q^2+M\frac{k^2}{\sqrt z}]dt+Q\dot z +\sigma
\ee
Hence the total boundary term can be expressed in the same form $\Sigma$, given in (25). So everything looks nice. The definition of the auxiliary variable and classical field equations may be realized as usual from the Euler-Lagrange equation, and the Hamiltonian is obtained as
\be
H=-p_{Q}p_{z}-M\frac{k}{z\sqrt z}p_{Q}^2+\frac{\sqrt z}{4M}Q^2-M\frac{k^2}{\sqrt z}.
\ee
For canonical quantization we express the Hamiltonian in terms of the basic variables as before, for which we choose $\dot z = x$ and replace $Q$ by $p_{x}$ and $p_{Q}$ by $-x$ and as a result obtain,
\be
H=xp_{z}+\frac{\sqrt z}{4M}p_{x}^2-M\frac{k}{z\sqrt z}x^2-M\frac{k^2}{\sqrt z}.
\ee
Finally the quantum dynamical equation reads,
\be
i\hbar \frac{1}{\sqrt z}\frac{\partial\psi}{\partial z}=-\frac{\hbar ^2 }{4Mx}\frac{\partial^2 \psi}{\partial x^2}-\frac{Mk}{ z^2 x}(x^2+kz)\psi.
\ee
Here again we do not require to consider any factor ordering, since $x$ and $p_{z}$ appearing in the Hamiltonian (33) are independent variables, likewise, $z$ and $p_{x}$. With a change of variable, $\alpha=\frac{2}{3}z^{\frac{3}{2}}$, the above equation can be expressed in a much convenient form,
\be
  i\hbar \frac{\partial\psi}{\partial \alpha}=-\frac{\hbar ^2}{4M x}\frac{\partial^2 \psi}{\partial x^2}-\frac{Mk}{ (\frac{3\alpha}{2})^\frac{4}{3} x}[x^2+k(\frac{3\alpha}{2})^\frac{2}{3}]\psi,
\ee
which can further be expressed as before,
\[ i\hbar \frac{\partial\psi}{\partial \alpha}=\hat{H_{0}}\psi\]

This time, we find that the effective Hamiltonian operator $\hat{H_{0}}$ is hermitian and hence there exists a standard quantum, mechanical probability interpretation of the theory as expatiated in our earlier works \cite{a:m}, \cite {a:s}. Further, we observe that we didn't have to choose an external time variable, rather an internal variable $\alpha$ acts as time parameter. We also notice that $\alpha = \frac{2}{3}z^{\frac{3}{2}}$ is proportianal to the proper volume $a^3$, which is naturally a resonably good choice for the time parameter. Finally, we end up with an effective time dependent potential,
\[V_{e}=-\frac{Mk}{x} (\frac{2}{3\alpha})^{\frac{4}{3}}[x^2+k(\frac{3\alpha}{2})^{\frac{2}{3}}],\]
 whose extremization leads to a power law inflationary solution of the scale factor
\[a=\sqrt{\frac{k}{2}}(t-t_{o}).\] 
Thus, we conclude that in the early universe, if Einstein-Hilbert action is modified by the introduction of $R^2$ term then power law inflation of the Universe is a natural quantum mechanical outcome, for which one does not require any additional field with a potential put in by hand. Further, it appears that graceful exit might follow naturally as the Universe transits from quantum to classical era if Einstein-Hilbert action would have been considered in addition. However, this is only a proposition that has not been shown explicitely in the present work. Hence, we rather conclude that {\large the action containing curvature squared term must be supplemented by a boundary term $\sigma$, given in equation (30)}. Let us now compare this, with our earlier work.
\section{\bf{Comparison with our earlier result}}
In our earlier works we have chosen variables different from that suggested by Boulware. Rather, we followed the work of Horowitz \cite {g:h} in the sense that instead of $Q^{ab}=\pi^{ab} = \frac{\partial A}{\partial ({\frac{\partial k_{ab}}{\partial t}})}$, we considered, $Q^{ab}=\pi^{ab} = \frac{\partial A}{\partial ({\frac{\partial \dot{h_{ab}}}{\partial t}})}$. Hence we did not require to make any change of variable and therefore, instead of (10) we started with action (7) and got rid of all removable total derivative terms to obtain,
\be
A=m\int~[a\ddot a^2+\frac{(\dot a^2+k)^2}{a}]dt+m(\frac{2}{3}\dot a^3 +2k\dot a),
\ee
where $m=72\pi ^2 \beta$. It should be noted that, as a result of removing all the removable total derivative terms, the action now has been supplemented by a boundary term $m(\frac{2}{3}\dot a^3 +2k\dot a)$, which is different from $\sigma$ given in (30) by an additive factor of $\frac{2}{3} m\dot a^3 $. This is definitely awkward. A particular action must be supplemented by the same boundary term, independent of the choice of variable. So, there is something wrong with our earlier proposal that all removable total derivative terms should be removed prior to the introduction of an auxiliary variable. However, as already stated, if we now
choose the auxiliary variable as $Q = \frac{\partial A}{\partial \ddot a} = 2m a\ddot a$ and express the action in canonical form, then
\[A=\int[-\dot Q \dot a+m\frac{(\dot a^2 +k)^2}{a}-\frac{Q^2}{4ma}]dt +Q\dot a+m(\frac{2}{3}\dot a^3 +2k\dot a).\]
It can easily be checked that classical field equations remain unchanged and the definition of auxiliary variable is recovered. However, the corresponding quantum cosmological equation changes considerably. It is now 
\be
i\hbar \frac{\partial\psi}{\partial \alpha}=\frac{\hbar ^2}{4mx}(\frac{\partial^2 \psi}{\partial x^2}+\frac{n}{x}\frac{\partial \psi}{\partial x})+\frac{m}{x}(x^2+k)^2 \psi,
\ee
where, $\alpha = ln~ a$ and $x=\dot a$, which is clearly different from that given in (35). However, this equation can again be expressed in the form,
 \be
i\hbar \frac{\partial \psi}{\partial \alpha}=\hat{H_{0}}\psi.
\ee
The effective Hamiltonian $\hat{H_{0}}$ is hermitian. Further, an effective potential has been evolved whose extremization leads either to vacuum Einstein's equation or inflation. This nice result that has been produced earlier \cite{a:m}, suggests that action corresponding to the curvature squared term should be supplemented by some boundary term $m(\frac{2}{3}\dot a^3 +2k\dot a)$ which is different from $\sigma$ given in (30). As a result, we finally end up with a boundary term  that contains $-96 \pi^2 \beta \dot a^3$ in addition to that given in equation (25). This result is not convincing. We expect that the final boundary term should be the same whether or not the action is supplemented with some boundary term ininitially. It is now clear that for different choice of variables we may have different removable total derivative terms in the action. In particular we have noticed that under the choice $z=\frac{a^2}{2}$, the removable total derivative term $2m\dot a^2 \ddot a$ appearing in the action (7) gets hidden in action (10). So our earlier proposal that all removable total derivative terms should be eliminated from the action is clearly wrong. In addition, we observe that the intrinsic time parameter here, is proportional to the natural log of the scale factor, rather than the proper volume as obtained in section (5), and which was naturally a better choice for such parameter. These are mainly the two drawbacks of our earlier works. This does not mean that one can not work with the set variables \{$h_{ab}, Q^{ab}=\pi^{ab} = \frac{\partial A}{\partial ({\frac{\partial \dot{h_{ab}}}{\partial t}})}$\}. Rather this suggests that the action (1) must be supplemented by a definite and appropriate boundary term. Hence, we withdraw our earlier proposal that all removable total derivative terms should be eliminated prior to the introduction of auxiliary variable and now suggest that { \large the action (1) must be supplemented by an appropriate
 boundary term $\sigma$, given by (30)}. To test our suggested prescription we go over to anisotropic and homogeneous cosmological models. 

\section{\bf{Problem arising in anisotropic models}}
 Homogeneous and anisotropic Kantowski-Sachs, axially symmetric Bianchi-1 and Bianchi-111 minisuperspace models have already been expressed altogether in equation (15), along with the expressions for Ricci scalar and extrinsic curvature given by equation (16).
In view of the definition of $Q^{ab}$, it is clear that for the model under consideration, a pair of auxiliary variables should be introduced. If we start with Boulware's prescription and don't remove total derivative terms, then the action (19) can be expressed in canonical form in view of the auxiliary variables (20) as 

\be
A=\int~[-\dot Q\dot x-\dot q\dot y-\frac{1}{8M}(\frac{Q^2 x^{\frac{3}{2}}}{y}+\frac{q^2 y}{4\sqrt x})-\frac{1}{2}\{\frac{1}{2}(\frac{\dot x}{x}-\frac{\dot y}{y})^2-\frac{k}{y}\}(Qx+\frac{1}{2}qy)]dt ~~+Q\dot x+q\dot y.
\ee
The boundary term $Q\dot x+q\dot y$ takes the form specified in (25) in view of the definitions of the auxiliary variables given in (20). Now, one can find $\frac{\partial L}{\partial Q}, \frac{\partial L}{\partial \dot Q}$ and $\frac{\partial L}{\partial q}, \frac{\partial L}{\partial \dot q}$, which are supposed to reproduce the definitions of the auxiliary variables $Q$ and $q$. Thus,
\[Q=4M\frac{y}{\sqrt x}[\frac{\ddot x}{x}-\frac{1}{2}\{\frac{1}{2}(\frac{\dot x}{x}-\frac{\dot y}{y})^2 -\frac{k}{y}\}],\]
and
\be
 q=16M\sqrt x[\frac{\ddot y}{y}-\frac{1}{4}\{\frac{1}{2}(\frac{\dot x}{x}-\frac{\dot y}{y})^2 -\frac{k}{y}\}].
\ee
 It shows that the definitions of the auxiliary variables $Q$ and $q$ given in (18) are not recovered and as a result, classical field equations will turn out to be wrong, although the action (39) is canonical. This implies that, $Q$ and $q$ can not be treated as independent auxiliary variables. This suggests that - {\large only one auxiliary variable suffices to express the action in canonical form.} Instead of (19), if one removes the removable total derivative terms from the action (19), and hence work with action (21), then this action can again be expressed in canonical form, in view of the auxiliary variables (22). However, the result is the same, ie., not the same auxiliary variables (22) are realized from the Euler-Lagrange equation and hence classical field equations again turn out to be wrong.  
\par
Same result is realized if one works even with Horowitz's prescription, ie if one considers $h_{ab}, \dot h_{ab}$ ie. $a, \dot a$ and $b, \dot b$ as the basic variables instead of $h_{ab}, k_{ab}$. In view of the definition of $^4 R$ given in (16), the action (1) can be expressed as,
  \[
A=16\pi\beta\int~~ 
{\Huge[}(\frac{b^2\ddot a^2}{a}+4a\ddot b^2+4b\ddot a\ddot 
b+4\frac{b}{a}\dot a\dot b\ddot a+2\dot b^2\ddot a+8\dot a\dot 
b\ddot b+4\frac{a}{b}\dot b^2\ddot b+2k\ddot a+4k\frac{a}{b}\ddot b+\]
\be
4\frac{\dot 
a^2\dot b^2}{a}+4\frac{\dot a\dot b^3}{b}+4k\frac{\dot a\dot 
b}{b}+\frac{a\dot b^4}{b^2}+2k\frac{a\dot 
b^2}{b^2}+k^2\frac{a}{b^2}){\Huge]}dt.
\ee                                                                       
As it is apparent, the lagrangian corresponding to the above action (41) contains both removable and unremovable second derivative terms. As par our earlier prescription let us first eliminate all the removable total derivative terms. In the present context of anisotropic and homogeneous minisuperspace models, completely removable terms appearing in the action (41) are $2k\ddot{a}, 4k\frac{a}{b}\ddot{b}$ and $ 4\frac{a}{b}\dot{b}^2\ddot{b}$. Further, $\dot{b}^2 \ddot{a}$ can be expressed in terms of $\dot{a}\dot{b}\ddot{b}$ and vice-versa. It has been observed that if one removes all these terms mentioned, and expresses $\dot{b}^2\ddot{a}$ in terms of $\dot{a}\dot{b}\ddot{b}$ or the other way round  and introduces two auxiliary variables $Q_{1} =-\frac{\partial{A}}{\partial{\ddot{a}}}$ and $q_{1} =-\frac{\partial{A}}{\partial{\ddot{b}}}$, the action can be expressed in canonical form but the definitions of the auxiliary variables are not recovered from the Euler-lagrange equation resulting in wrong classical field equations. Actually it does not require to express $\dot{b}^2\ddot{a}$ in terms of $\dot{a}\dot{b}\ddot{b}$ or the other way round as par our earlier proposal, since these are not completely removable total derivative terms. So next, if one removes only the three completely removable total derivative terms keeping the interchangeable terms intact and follow the same procedure, then it is observed that the action can not be expressed in canonical form. The reason has already been stated above, ie. the two auxiliary variables are not independent. 
\par
In view of the results that have been discussed above we disregard Ostrogrdski and Boulware's prescription of introducing as many auxiliary variables as the number of metric co-efficients and proceed in the following section introducing only a single auxiliary variable.
\section{\bf{One auxiliary variable}} 
As, already discussed, we now introduce only one auxiliary variable to cast the action (41) in canonical form and to obtain the correct classical field equations. Here we follow our earlier proposal to eliminate all the removable total derivative terms from the action prior to the introduction of auxiliary variable. So interchangeable terms $\dot{b}^2\ddot{a}$ and $\dot{a}\dot{b}\ddot{b}$ are kept intact and after the removal of all the three removable total derivative terms viz., $2k\ddot{a}, 4k\frac{a}{b}\ddot{b}$ and $ 4\frac{a}{b}\dot{b}^2\ddot{b}$ from the action (41), we are left with, 
\[
A=16\pi\beta\int~~ 
[(\frac{b^2\ddot a^2}{a}+4a\ddot b^2+4b\ddot a\ddot 
b+4\frac{b}{a}\dot a\dot b\ddot a+2\dot b^2\ddot a+8\dot a\dot 
b\ddot b+\]
\be
4\frac{\dot 
a^2\dot b^2}{a}+\frac{8\dot a\dot b^3}{3b}+\frac{7a\dot b^4}{3b^2}+6k\frac{a\dot 
b^2}{b^2}+k^2\frac{a}{b^2})]dt+16\pi\beta[2k(\dot a +2\frac{2a\dot b}{b})+\frac{4a\dot{b}^3}{3b}].
\ee 
It is to be noted that the boundary term $16\pi\beta[2k(\dot a +2\frac{a\dot 
b}{b})+\frac{4a\dot{b}^3}{3b}]$ is different from $\sigma$ given in (30) due to the presence of the last term viz. $16\pi\beta\frac{4a\dot{b}^3}{3b}$.
Now, let us introduce an auxiliary variable as
\be
Q_{1}=\frac{\partial{A}}{\partial{\ddot{a}}}=16 \pi \beta [2\frac{b^2}{a}\ddot{a}+4b\ddot{b}+4\frac{b}{a}\dot{a}\dot{b}+2\dot{b}^2].
\ee
In view of the above variable $Q_{1}$ one can express the action (42) in the following canonical form, with $M_{1}=16\pi\beta$, 
\be
A=\int[(\ddot a+2\frac{a}{b}\ddot b) Q_{1}-\frac{a Q_{1}^2}{4M_{1} b^2}+\frac{Q_{1} a}{b^2}(2\frac{b}{a}\dot a\dot b+\dot b^2)+6ka\frac{\dot b^2}{b^2}+k^2\frac{a}{b^2}],
\ee
provided, the action (42) contains an additional term,  viz., $4\frac{a}{b}\dot{b}^2 \ddot{b}$. Notice that, this term is a total derivative term that was present in action (41) and so was eliminated from the action at the begining along with two other terms. This term contributed  $16\pi\beta\frac{4a\dot{b}^3}{3b}$ to the surface term for which the surface term could not be expressed in the form given by $\sigma$ in (30). It is further noticed that, this term does not contribute to the auxiliary variable (43). Hence, we understand that not all total derivative terms should be eliminated from the action as proposed in our earlier works \cite{a:m}, \cite{a:s}, \cite{a:n}. The boundary term in (42) is now $M_{1}[2k(\dot a +2\frac{a\dot b}{b})]$, which can  be expressed exactly in the form given by $\sigma$ in (30). So, we notice that in view of a single auxiliary variable canonization of the action (41) now forces us to retain required total derivative terms in the action. So, anisotropic models confirm our proposition that action (1) must be supplemented by a boundary term $\sigma$ having the form given in (30). 
\par
In view of this result we understand that a boundary term $\sigma$ should supplement the action (41), so let us write the correct steps to avoid confusion, if any. Action (42) should be replaced by, 
\[
A=M_{1}\int~~ 
[(\frac{b^2\ddot a^2}{a}+4a\ddot b^2+4b\ddot a\ddot 
b+4\frac{b}{a}\dot a\dot b\ddot a+2\dot b^2\ddot a+8\dot a\dot 
b\ddot b+4\frac{a}{b}\dot{b}^2\ddot{b}\]
\be
4\frac{\dot 
a^2\dot b^2}{a}+4\frac{\dot a\dot b^3}{b}+\frac{a\dot b^4}{b^2}+6k\frac{a\dot 
b^2}{b^2}+k^2\frac{a}{b^2})]dt+M_{1}[2k(\dot a +2\frac{a\dot 
b}{b})]].
\ee 
The auxiliary variable $Q_{1}$ given by (43) remains unchanged, and the action (45) can now be expressed in the canonical form along with the final boundary term as, 
\be
A=M_{1}\int~~[-\dot Q_{1}(\dot a+2\frac{a}{b}\dot b)+3\frac{a\dot b^2}{b^2}Q_{1}-\frac{aQ_{1}^2}{4b^2}+\frac{ka}{b^2}(6\dot b^2+k)]dt+(2kM_{1}+Q_{1})(\dot a+2\frac{a}{b}\dot b)
\ee
The final boundary term $(2kM_{1}+Q_{1})(\dot a+2\frac{a}{b}\dot b)$, appearing in (46) can be cast exactly in the form given in (25). So everything looks nice, the definition of $Q_{1}$ is recovered, so also the correct classical field equations. However, this choice of basic configuration space variables viz., $h_{ab}, \dot h_{ab}$ ie. $a, \dot a, b$ for which the auxiliary variable is $Q_{1}=\frac{\partial{A}}{\partial{\ddot{a}}}$ (or even if we choose $a, b, \dot b$ for which the auxiliary variable is $q_{1}=\frac{\partial{A}}{\partial{\ddot{b}}}$)  put up a problem during quantization. The Hamiltonian in the present situation is, 
\be
H=-[\frac{1}{M_{1}}p_{a}p_{Q_{1}}-\frac{b^2}{12 M_{1} a(Q_{1}+2k)}[p_{b^2}+4\frac{a^2}{b^2}p_{a}^2-4\frac{a}{b}p_{a}p_{b}]-\frac{M_{1}a}{4 b^2}(Q_{1}^2-4k^2)].
\ee

Now, our basic variables are ($a, \dot a, b, p_{a}, p_{\dot a}, p_{b}$), instead of which we worked with ($a, Q_{1} (= p_{\dot a}), b, p_{a}, p_{Q}, p_{b}$). So, for quantization through basic variables let us choose $\dot a = x$, hence replace $Q_{1}$ by $p_{x}$, but since $p_{Q_{1}}= -(\dot a +2\frac{a}{b}\dot b)$, therefore it can not be replaced simply by $-x$, rather it involves some more terms involving momenta $p_{a}$ and $p_{b}$. Therefore we conclude that this variable is not healthy for quantization of the theory. 
\par
Instead of $Q_{1}=\frac{\partial{A}}{\partial{\ddot{a}}}$, one can even choose $q_{1}=\frac{\partial{A}}{\partial{\ddot{b}}}$, as the auxiliary variable, which in view of action (42) is given by,
\[q_{1}= 4b\ddot a+8a\ddot b+8\dot a\dot b.\]
Hence, action (42) can be expressed in the following canonical form,
\[A=\int[-\dot q_{1}(\frac{b\dot a}{2a}+\dot b)-\frac{q_{1}^2}{16 a}+q_{1}(\frac{\dot b^2}{2b}+\frac{\dot a \dot b}{2a}+\frac{b\dot a^2}{2a^2})+ q_{1} independent~ terms] + \Sigma_{1}.\]
Not only that $\Sigma_{1}$ obtained in the process is different from that given in (25), but also the auxiliary variable obtained from the above canonical action, in view of Euler-Lagrange equation 
\[q_{1}= 4b\ddot a+8a\ddot b+8\dot a\dot b+4\frac{a}{b}\dot b^2,\] 
is clearly different from one we started with. Thus, it will again yield wrong classical field equations. This result is unique and astonishing, since total derivative term now directly influences classical field equations. In order to recover the same $q_{1}$ from the canonical action and as such to obtain the correct classical field equation,s one has to keep the total derivative term  $4\frac{a}{b}\dot{b}^2 \ddot{b}$ in the action, right from the begining. This is for the simple reason that the term  $4\frac{a}{b}\dot{b}^2 \ddot{b}$, that did not contribute to  $Q_{1}$, now contributes to the auxiliary variable $q_{1}$. This again proves our proposition that action (1) must be supplemented by a boundary term $\sigma$ given by (30), otherwise classical field equations might turn out to be wrong. The problem of quntization of the theory is not resolved with this variable too, for the same reason discussed above. However, in view of such wonderful result, we are now force to conclude that {\large total derivative terms play a crucial role in the classical domain and not all canonical actions are liable to produce correct classical field equations}.   
\par
It appeared in the context of isotropic model that $z = \frac{a^2}{2}$ is a better variable, since, in view of this variable the removable total derivative terms that appeares in the action (13) is exactly $\sigma$, given in (30). But this essentially due to the simplicity of isotropic model and is not true in general. In the anisotropic model with $x = \frac{a^2}{2}, y = \frac{y^2}{2}$, the removable total derivative terms in the action (21) is not simply $\sigma$ given in (30) but it contains some additional terms which is apparent from (21). However, if the action is supplemented by $\sigma$ given in (30) then it can be cast in the following canonical form in view of a single auxiliary variable $Q_{2}=\frac{\partial A}{\partial \ddot x}$, (say), and with $M=8\sqrt 2 \pi \beta$, 
\[
A=\int~~[-\dot Q(\dot x+2x\frac{\dot y}{y})-Q\frac{x}{2}(\frac{\dot 
x^2}{x^2}-3\frac{\dot y^2}{y^2}+2\frac{\dot x\dot 
y}{xy}+2\frac{k}{y})-\frac{Q^2 x^\frac{3}{2}}{4My}]dt~~+Q(\dot 
x+2x\frac{\dot y}{y}). \]
The definition of $Q$ is recovered and correct classical field equations are obtained from the Euler-Lagrange equations. Further, the same boundary term, given in (25) is found. However, the same problem arises during quantization, since, $p_{Q}=-\dot x-2x\frac{\dot y}{y}$, therefore, quantization through basic variables can not be performed. 
\par
At this stage, since we have been able to convince ourselves that an action containing curvature squared term must be supplemented by an appropriate boundary term given by $\sigma$ in (30), it might appear that a pair of auxiliary variable might also yield correct classical field equations provided the action is supplemented by the said boundary term. To check this, we take action (45) and introduce a pair of auxiliary variables 
\[Q_{1}=\frac{\partial{A}}{\partial{\ddot{a}}}=16 \pi \beta [2\frac{b^2}{a}\ddot{a}+4b\ddot{b}+4\frac{b}{a}\dot{a}\dot{b}+2\dot{b}^2]\]
and 
\[q_{1}=\frac{\partial{A}}{\partial{\ddot{b}}}= 4b\ddot a+8a\ddot b+8\dot a\dot b+4\frac{a}{b}\dot b^2. \]
Hence action (45) can now be expressed in the following canonical form,
\[A=\int[-\dot Q \dot a-\dot q \dot b -\frac{aQ^2}{8b^2}+Q(\frac{a\dot b^2}{2b^2}+\frac{\dot a \dot b}{2a})+6k\frac{\dot a\dot b^2}{b^2}+\frac{ak^2}{b^2}]dt+\Sigma.\]
In view of the above canonical action, it is readily found that same auxiliary variables are not recovered, since they are now,
\[Q_{1}=4\frac{b^2}{a}\ddot a +2\dot b^2 +4\frac{b}{a}\dot a\dot b\]
and,
\[q_{1}=16 a\ddot b +4\frac{a}{b}\dot b^2 +8\dot a\dot b,\]
which results in wrong classical field equations. Hence Ostrogradsi's prescription \cite{o:m}, Boulware's proposal \cite {b:et} and Horowitz's \cite{g:h} suggestions should be out and out rejected, since all of them suggested to consider the same number of auxiliary variables as the number of metric coefficients. Therefore, we conclude that to obtain correct classical field equations one has to consider only one auxiliary variable and the action must be supplemented by an appropriate boundary term prior to the introduction of the auxiliary variable.
\par
Now, so far as the classical field equations are concern, we find that all set of variables work as well. However, for canonical quantization of the theory none of them is found suitable. Thus, Ostrogradski's prescription \cite{o:m} of considering the phase space variables  $h_{ab}$, a quantity $Q_{ab}$ realated to the extrinsic curvature tensor and their conjugate momenta, $p^{ab}, P^{ab}$, for canonical quantization of higher order theory of gravity, is unacceptable in general and so is  Boulware's prescription \cite{b:et}. Hence, for the anisotropic minisuperspace models under consideration, we need to search for a variable such that the action expressed in terms of that variable contains $\sigma$ as the only removable total derivative term along with the possibility of having only a single auxiliary variable.
\par
For this purpose, let us consider a variable $z = ab^2$, in view of which, the Ricci scalar (16) takes the form
\be
^4 R = 2(\frac{\ddot z}{z}-2\frac{\dot z \dot b}{z b}+3\frac{\dot b^2}{b^2}+\frac{k}{b^2}).
\ee
Hence, the action (1) is 
\be
A=M\int~[\frac{\ddot z^2}{z}-4\ddot z\frac{\dot z\dot b }{zb}+6\ddot z\frac{\dot b^2}{b^2}+4\frac{\dot z^2 \dot b^2}{zb^2}-12\dot z\frac{\dot b^3}{b^3}+9z\frac{\dot b^4}{b^4}+6kz\frac{\dot b^2}{b^4}+\frac{k^2 z}{b^4}]dt+2M \dot z\frac{k}{b^2},
\ee
where, $M = 16\pi \beta$. It is important to note that, now we are left with only one removable total derivative term, $2M \dot z\frac{k}{b^2}$, can be expressed exactly in the form given by $\sigma$ in (30). Further there do not arise any confusion regarding interchangeable terms because they are now absent from the above action (49). Under such a choice of variable we can inaddition have only one auxiliary variable
\be
Q=\frac{\partial A}{\partial \ddot z} = 2M [\frac{\ddot z}{z}-2\frac{\dot z \dot b}{z b}+ 3\frac{\dot b^2}{b^2}].
\ee
Hence, all of our requirements are thus fulfilled. The above action (49) can now be expressed in the following canonical form 
\be
A=\int~[-\dot Q \dot z-\frac{Q^2 z}{4M}-2Q\dot z\frac{\dot b}{b}+3Q z\frac{\dot b^2}{b^2}+M\frac{k}{b^2} z(6\frac{\dot b^2}{b^2}+\frac{k}{b^2})]dt+Q\dot z+\sigma
\ee
The final boundary term now takes the form $\Sigma$ given in (25). In view of the above form of the canonical action (51), one can recover the definition of $Q$ and find the correct classical field equations. However, the Hamiltonian now takes the form,
\be
H=p_{z}p_{Q}+z\frac{Q^2}{4M}+\frac{b^2(bp_{b}-2Qp_{Q})(bp_{b}+6Qp_{Q})}{12z(b^2 Q+2Mk)}-M\frac{k^2 z}{b^4}.
\ee
Now under the choice $\dot z = u$, we can now quantize the theory with basic variables, by replacing $Q$ by $p_{u}$ and $p_{Q}$ by $-u$. In view of such replacements the Hamiltonian (52) can be expressed as,
\be
H=-up_{z}+\frac{z}{4M}p_{u}^2 +\frac{b^2(bp_{b}+2up_{u})(bp_{b}-6u_{p_{u}})}{12 z (b^2 p_{u}+2Mk)}-4z \frac{k^2}{b^4}
\ee
The Hamiltonian (53) is highly complicated due to the presence of momentum $p_{u}$ in the denominator of the third term. Although, under quantization, $z=ab^2$, which is essentially the proper volume, acts as an internal time parameter, as in isotropic model, it does not produce an effective Hamiltonian that is hermitian. However, the trouble has originated due to the presence of the two middle terms in the Ricci scalar (48). If one can find a variable that combines the two terms, then the complicacy of the Hamiltonian (53) will disappear, and the effective Hamiltonian might turn out to be hermitian. We have neither been able to find such a variable nor been able to prove if it exists. Nothing to worry about, even if it does not exist, since nonhermiticity of the effective Hamiltonian corresponding to curvature squared action might be a characteristic feature of any anisotropic model and as a result a quantum mechanical probability interpretation is not possible. It has been mentioned in the introduction, that unitarity of higher derivative quantum theory of gravity has been established by Tomboulis \cite{t:p}. So, it might be conjectured along the same line that, unitarity of the effective Hamiltonian will be restored if the most general form of curvature squared action is taken into account. 

\section{\bf{Concluding remarks}}
In connection with Hamiltonian formulation of an action containing higher order curvature invariant terms, - curvature squared term in particular, in the present context, in both isotropic and anisotropic (ofcourse homogeneous) minisuperspace models, a few important results have emerged, one of which might shatter our age old idea. In the following we list the set of results.
\par
Firstly, for canonical quantization of such a theory the whole superspace is usually spanned by the phase space variables $h_{ab}, Q_{ab}, p^{ab}, P^{ab}$, according to the suggestion of Ostrogrdski \cite{o:m}, where
\[ Q^{ab} = 2k_{ab}\]
and
\[P_{ab}= -\beta\sqrt{h}h^{ab}~R\]
in the absence of the electric part of the Weyl tensor, and $k_{ab}$ is the extrinsic curvature tensor. It has been shown in section (2) that such a definition of momentum is not at all reliable, since it does not tally with the usual canonical definition of momentum. As a result, Hamiltonian constraint equation in terms of the phase space variables and hence the quantum dynamical equation is clearly wrong. Consideration of the electric part of the Weyl tensor in the definition of momentum does not change the situation, since, corresponding to the curvature squared action under consideration the canonical definition of momentum remains unchanged. 
\par
Next, we have proved the need for a boundary term for the quantum formulation of such a theory and finally been able to show that correct Hamiltonian formulation of an action containing curvature squared term in particular, requires that the action should be supplemented by a boundary term  $\sigma=\int ~^{3}R~ K~ \sqrt{h}~d^{3}x$. 
\par
Ostrogrdski's prescription \cite{o:m} suggests, there should be as many auxiliary variable as the number of metric coefficients. It has been shown in connection with a set anisotropic models that this is not true. It is not possible to find correct classical field equations unless only a single auxiliary variable is chosen. Therefore, Ostrogradski's prescription \cite{o:m} along with Boulware's proposal \cite{b:et} are vulnarable. It has been further shown that, once the action is supplemented by the correct boundary term $\sigma$, any one of the auxiliary variables can produce correct classical field equation. However, if the action is not supplemented by the appropriate boundary term $\sigma$, it might produce wrong classical field equation. This is a rather shocking result, since this goes against our age old conviction that total derivative terms do not contribute to classical field equations. So we end up with two specific conditions, viz., {\large not more than one auxiliary variable should be taken into account and action containing curvature squared term must be supplemented by an appropriate boundary term $\sigma$}. Further, we understand that total derivative terms play crucial role even in classical domain and a canonical action do not always guarantee correct classical field equations. 
\par
It has already been mentioned that if the two conditions are satisfied then any of the auxiliary variables is capable of producing correct classical field equations. However it has been shown that these variables are incapable of producing correct quantum dynamical equation. To obtain correct quantum dynamical equation corresponding to such an action, one has to find a variable such that when the action is expressed in terms of that variable, it must not contain any total derivative term other than $\sigma$. Further, there should exist the possibility of finding only one auxiliary variable. This variable has been found to be the proper volume in both isotropic and anisotropic models that have been considered and it acts as the internal time parameter in the corresponding quantum dynamical equation. Hence, there exists a possibility of casting a general quantum dynamical equation corresponding to an action containing curvature invariant term, which will be considered in a future article. The effective Hamiltonian appearing in the anisotropic model suffers from the disease of nonunitarity. This might get removed when an action containing most general form of higher order curvature invariant terms will be considered.    
\par
Finally, let us point out the whole scheme of canonical quantization of the curvature squared action, 
\[A=\beta\int~{d^4x}\sqrt{-g}~R^2.\] 
1.Supplement the action by a boundary term,
\[\sigma=4\beta\int~K~\sqrt h~^3 R~d^3 x.\]
2.Replace one of the geometric variables by the proper volume $z$ (say), and define an auxiliary variable, 
\[Q=\frac{\partial A}{\partial \ddot z} = p_{\dot z}.\]
3.Express the action in canonical form in terms of the auxiliary variable $Q$ and other field variables, and get rid of the excess boundary term. Thus the total boundary term is now 
\[\Sigma=4\beta\int~K~\sqrt h~^4 R~d^3 x.\]
 Now, one obtains correct classical field equations and Hamiltonian formulation of the theory is thus performed.
\par
4.Finally, canonical quantization is performed in terms of the basic variables, which are $x=\dot z$ and $p_{x}$ along with other geometric field variables and their conjugate momenta. For this purpose, replace $Q$ by $p_{x}$ and $p_{Q}$ by $x$. Thus a schr\"{o}dinger like equation is revealed.

\end{document}